\documentclass[reprint,aps]{revtex4-1}
\usepackage[british]{babel}
\usepackage{amsmath,amssymb,graphicx,xcolor,theorem}

\newcommand{\nobracket}{}
\newcommand{\tmcolor}[2]{{\color{#1}{#2}}}
\newcommand{\tmmathbf}[1]{\ensuremath{\boldsymbol{#1}}}
\newcommand{\tmop}[1]{\ensuremath{\operatorname{#1}}}

\newcommand{\tmtextit}[1]{\text{{\itshape{#1}}}}

\newcounter{nnacknowledgments}

{\theorembodyfont{\rmfamily}\newtheorem{acknowledgments*}[nnacknowledgments]{Acknowledgments}}
\begin{document}

\title{Macroscopic QED and noise currents in time--varying media}

\author{S. A. R. Horsley$^1$ and R. K. Baker$^1$}
\affiliation{$^1$School of Physics and Astronomy,\\
University of Exeter,\\
Stocker Road,\\
Exeter, UK\\
EX4 4QL}

\

\begin{abstract}
  Macroscopic QED (MQED) is the field theory for computing quantum electromagnetic effects in dispersive media.  Here extend MQED to treat \emph{time--varying}, dispersive media.  For a time dependent Drude model, we find that the expected replacement $\varepsilon (\omega)\rightarrow \varepsilon (t, \omega)$ within standard MQED leads to nonphysical polarization currents, becoming singular in the limit of a step change in carrier density.  We show this singular behaviour can be removed through modifying the reservoir dynamics, quantizing the resulting theory and finding the non--equilibrium, time--varying noise currents, which exhibit extra correlations due to temporal reflections within the material dynamics.
\end{abstract}

{\maketitle}

The arrow of time is evident in most electromagnetic (EM) problems.  Sources radiate \emph{outward} going waves, and matter responds only to EM fields in the \emph{past}.  By contrast, there is no such ``arrow of
space''.  This inequivalence of space and time implies that waves with a \emph{time} dependent speed, $c (t)$ can't be understood through a simple
$x \leftrightarrow t$ relabelling of waves with a \emph{space} dependent
speed, $c(x)$. For instance, an abrupt change of wave speed within a region
of space causes reflection; the familiar conversion between waves propagating towards $x = + \infty$ and $- \infty$.  A similar abrupt change in time does \emph{not}, by analogy, generate waves that propagate into the past~\cite{bacot2016,moussa2023}.  That would violate causality.  This
inequivalence is especially conspicuous in quantum electrodynamics (QED), where, unlike reflection from a spatial boundary, temporal reflection implies the creation of photons~\cite{mendonca2005,dezael2010}.

Here we examine the theory of QED in a dispersive, time--varying material.  We consider materials described by a time--dependent Drude model, although our conclusions can be applied more generally.  Our first task is to find a classical Lagrangian (and by extension, a Hamiltonian) that accurately describes the field and material dynamics.  This is a difficult task as the presence of dispersion implies dissipation, which---as is well known in the theory of macroscopic QED (MQED)---forces us to introduce a reservoir of simple harmonic oscillators to mimic the material dynamics.  We examine two possible modifications to the field--reservoir Lagrangian to describe a time modulated material response, finding that modulating the field--reservoir coupling leads to an unphysical permittivity and anomalously large noise currents.

\begin{figure}[h]
  \includegraphics[width=7.84149940968123cm,height=6.97022169749443cm]{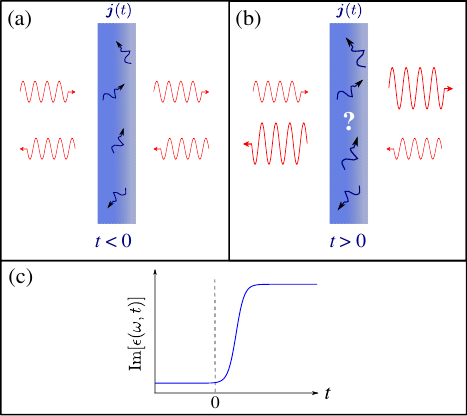}
  \caption{\textbf{Noise currents in time--varying media:} (a) In
  equilibrium, absorption of electromagnetic energy is balanced by
  re--emission, via fluctuating `noise currents', $\tmmathbf{j}$ within the
  material.  These can be calculated using MQED. (b) If the
  coupling between field and medium is changed abruptly as in (c), absorption and re--emission are no longer in balance and it is unclear how the noise currents should behave.  Here we examine extensions to MQED to treat this problem.\label{fig:schematic} \ }
\end{figure}

These fluctuating (noise) currents are a crucial prediction of MQED.  In stationary systems these noise currents obey the fluctuation--dissipation theorem (FDT)~\cite{kubo1966,volume9}, which derives from the equilibrium between absorption and emission (as in the schematic in Fig. \ref{fig:schematic}a).  As time--varying materials are, by definition \emph{not} in thermal equilibrium, the FDT does not apply.  We might expect an abrupt modulation of the material to radiate, due to the abrupt change in the noise currents (Fig. 1b,c).  The second part of our work finds the evolution of these fluctuating currents during a modulation of the material parameters.  Our findings complements the recent paper of V{\'a}zquez--Lozano et al. who investigated modifications to these currents using perturbation theory and the assumption of a non--dispersive
modulation of the material~\cite{lozano2023}.

Besides fundamental interest, the reason for paying such close attention to
the quantum theory of time--varying materials comes from recent experiments,
where large and abrupt changes in material parameters have been implemented at radio frequencies~\cite{moussa2023}, optical frequencies~\cite{zhou2020,bohn2021,lustig2023,tirole2023}, and for acoustic~\cite{chen2021,wen2022}, elastic~\cite{nassar2020}, and water waves~\cite{bacot2016}.  Note that dispersion is non--negligible in all of
these experiments.  In parallel, analytical and numerical calculations
have shown a number of new effects in these materials, such as synthetic
motion and drag~\cite{huidobro2019,huidobro2021}, temporal aiming~\cite{pena2020}, gain and wave compression~\cite{galiffi2021b,pendry2021b}, spectrum reshaping~\cite{zhang2018}, non--reciprocity~\cite{shaltout2015}, and have found special pulses with enhanced or suppressed transmission~\cite{horsley2023b}.  For a comprehensive review see~\cite{galiffi2022}.

More recently a number of authors have considered more subtle, quantum effects in time and space--time varying materials.  These papers are mostly
concerned with calculating the rate of photon creation~\cite{lamb2016,prain2017,sloan2022,horsley2023}, and with only a few exceptions, dispersive effects are usually not taken into account.

\section{Describing Dispersion}

In classical electromagnetism the susceptibility of a material is something we simply impose, making as lunatic a choice as we like.  In quantum
electromagnetism we are not so free.  As soon as material dispersion---and
by implication, dissipation~\cite{bohren2010}---is accounted for, the usual
electromagnetic Hamiltonian becomes non--Hermitian, breaking the conservation of norm, and equivalently the operator commutation relations.  The moral is that, to quantize a system properly, the Hamiltonian must include the material degrees of freedom.

MQED~\cite{huttner1992,scheel2008,philbin2010} represents the
solution to this problem: a general approach to the quantum mechanics of the
electromagnetic field in a dispersive, dissipative material.  The theory
can be seeded from different assumptions and here we work from a Lagrangian,
as in Ref.~\cite{philbin2010}.  The Lagrangian of MQED is given as the integral of the Lagrangian density, which is broken into three pieces,
\begin{equation}
    \mathcal{L} = \mathcal{L}_{\tmop{EM}} + \mathcal{L}_{\text{I}} +
    \mathcal{L}_R .\label{eq:Ldensity}
\end{equation}
For simplicity here we work in one spatial dimension plus time, where the
Lagrangian density for the electromagnetic field equals the textbook
expression {\cite{volume2}},
\begin{align}
    \mathcal{L}_{\tmop{EM}} & = \frac{\varepsilon_0}{2} [\varepsilon_{\infty}
    E^2 - c^2 B^2]\nonumber\\
    & = \frac{\varepsilon_0}{2} \left[ \varepsilon_{\infty} \left(
    \frac{\partial A}{\partial t} \right)^2 - c^2 \left( \frac{\partial
    A}{\partial x} \right)^2 \right],\label{eq:LEM}
\end{align}
where $\varepsilon_{\infty}$ is the high frequency limit of the permittivity, $A$ is the vector potential with the electric field given by $\boldsymbol{E} =-\partial_t A \boldsymbol{e}_y$, and the magnetic field by $\boldsymbol{B} = \partial_x A \tmmathbf{e}_z$.  Assuming the material responds linearly to the EM field, we add the Lagrangian density of a continuum of harmonic oscillators (``the reservoir''), with each oscillator having amplitude $X_{\omega}$,
\begin{equation}
    \mathcal{L}_R  = \frac{1}{2} \int_0^{\infty} \text{d} \omega \left[
    \left( \frac{\partial X_{\omega}}{\partial t} \right)^2 - \omega^2
    X_{\omega}^2 \right], \label{eq:LR}
\end{equation}
which, given the stupendous number of degrees of freedom, we hope might mimic the linear response of any material we choose.  Indeed, for stationary
media, this is the case.  The EM field and reservoir are coupled via the
following interaction
\begin{align}
    \mathcal{L}_{\text{I}} & = - \frac{\partial A}{\partial t} \int_0^{\infty}
    \text{d} \omega \alpha (\omega) X_{\omega}\nonumber\\
    & = - \frac{\partial A}{\partial t} \int_0^{\infty} \text{d} \omega
    \sqrt{\frac{2 \omega \varepsilon_0 \tmop{Im} [\varepsilon (\omega)]}{\pi}}
    X_{\omega}.\label{eq:LInt}
\end{align}
where $\varepsilon (\omega)$ is the complex permittivity, which must be purely dissipative (i.e. no gain, $\tmop{Im} [\varepsilon] > 0$) for our Lagrangian to be real valued.  Through applying Lagrange's equations of motion to the above sum of Eqns. (\ref{eq:LEM}--\ref{eq:LInt}) and eliminating the reservoir as
described in Ref.~\cite{philbin2010}, the EM field obeys Maxwell's
equations, with the effective permittivity given by,
\begin{equation}
  \varepsilon (\omega) = \varepsilon_{\infty} + \text{P} \frac{1}{\pi} \int_{-
  \infty}^{\infty} \frac{\tmop{Im} [\varepsilon (\omega')]}{\omega' - \omega}
  \text{d} \omega' + \text{i} \tmop{Im} [\varepsilon (\omega)] .
  \label{eq:effective-permittivity}
\end{equation}
The real part of this effective permittvity (\ref{eq:effective-permittivity}) is the Hilbert transform of the imaginary part, a relationship known as a Kramers--Kronig relation {\cite{bohren2010}}, which holds for any causal material response.  Provided we assume that our material is purely dissipative and causal, the equations of motion derived from the Lagrangian density (\ref{eq:Ldensity}) reproduces Maxwell's equations for any permittivity we like.  Extensions of this approach can be applied to mimic the response of magnetic~\cite{philbin2010}, bianisotropic~\cite{horsley2011}, and moving media~\cite{horsley2012}.  Versions of this theory have been used to calculate
dispersion forces~\cite{buhman2007,philbin2011}, spontaneous emission~\cite{feist2021}, and quantum friction~\cite{silveirinha2014,horsley2015}.  For a review see~\cite{scheel2008}.

\begin{figure}[h]
  \includegraphics[width=8.01575823166732cm,height=8.01575823166732cm]{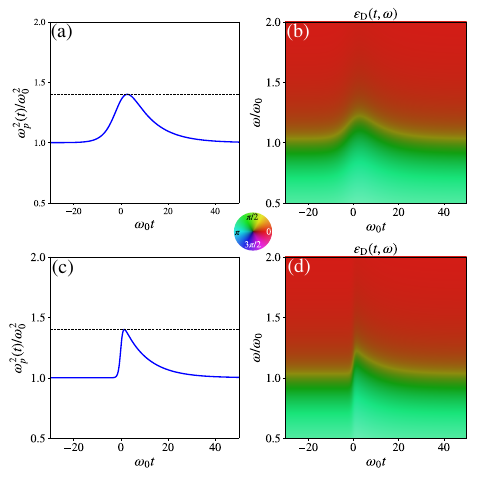}
  \caption{\textbf{Approximate Drude model:} Panels (c--d) show the assumed
  plasma frequency in Eq. (\ref{eq:epsilon-drude-approx}), which varies as a
  smoothed step function, $\omega_p^2 (t) = \omega_0^2 + \delta \omega_p^2
  \frac{1}{2} (1 + \tanh (t / t_r)) e^{- t / t_d}$.  In both
  cases $\delta \omega_p^2$ is chosen to give the same maximum plasma
  frequency $\omega_{\max}^2 = 1.4 \omega_0^2$ shown as the black dashed
  line.  Panels (a--b) show a plasma frequency changing slowly, $\omega_0
  t_r = 5$ and $\omega_0 t_d = 10$, whereas (c--d) considers a rapid change,
  $\omega_0 t_r = 1$ and $\omega_0 t_d = 10$.  Panels (b--d) show the
  complex permittivity for these two cases (colour shows phase and saturation magnitude, as indicated in the central panel).
  \label{fig:approximate-drude}}
\end{figure}

Now suppose the material is time--varying.  For example, the large and
rapid modulation of Indium--Tin--Oxide (ITO) reported in~\cite{zhou2020,bohn2021,lustig2023,tirole2023}, a material which is both
lossy and highly dispersive.  There are several ways to implement the time
variation in the Lagrangian (\ref{eq:Ldensity}).  We can include a time
dependence in the field--reservoir coupling (\ref{eq:LInt}), or make the
Lagrangian of the reservoir (\ref{eq:LR}) explicitly time dependent, or some
combination of the two.  The remainder of the paper explores the
consequences of these different choices.

\subsection{Special case: modulated metals}\label{sec:modulated-metals}

\begin{figure}[h]
  \includegraphics[width=8.01575823166732cm,height=8.01575823166732cm]{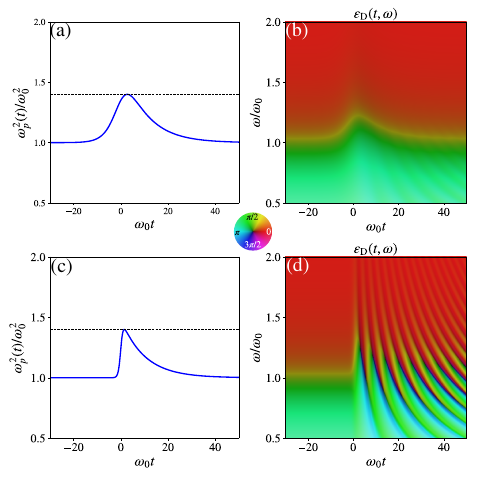}
  \caption{\textbf{Stepanov model:} Parameters identical to Fig.~\ref{fig:approximate-drude}, but with the permittivity equal to the Stepanov model defined in Eq. (\ref{eq:epsilon-drude}).  As anticipated, when the relative change in the plasma frequency is small (a--b), the permittivity of the Stepanov model approaches that of the approximate Drude model (\ref{eq:epsilon-drude-approx}), meanwhile for rapid changes (c--d) the frequency--time permittivity develops oscillations. \label{fig:stepanov}}
\end{figure}

For illustration we take a Drude model with a time--varying plasma
frequency.  This model is commonly used to model e.g. the EM response of
time modulated ITO, where the permittivity has been previously written as (see e.g. {\cite{bohn2021,tirole2022}}),
\begin{equation}
  \varepsilon_{\text{D}} (t, \omega) \sim \varepsilon_{\infty} -
  \frac{\omega_{\text{p}}^2 (t)}{\omega \left( \omega + \text{i} \gamma
  \right)}, \label{eq:epsilon-drude-approx}
\end{equation}
with $\gamma$ the collision rate and $\omega_p^2 = e^2 N / \varepsilon_0
m^{\ast}$ the plasma frequency, written in terms of the electron charge $e$,
effective mass $m^{\ast}$, and carrier density $N$.  Although not widely
stated, Eq. (\ref{eq:epsilon-drude-approx}) is an approximation to the
true permittivity, valid for a slowly modulated material response (Fig.
\ref{fig:approximate-drude} shows the behaviour of the permittivity
(\ref{eq:epsilon-drude-approx}) for adiabatic and non--adiabatic modulations)
.
Here we assume the modulation of the plasma frequency is due to a changing
carrier density.  An almost identical argument can be made for a fixed
density of carriers with a time varying effective mass, which is most relevant for ITO {\cite{bohn2021}}.  But through assuming a time varying carrier density we can directly compare our results to Stepanov's earlier work on time--varying plasmas~\cite{stepanov1976}.  For us, the important
finding in Ref. {\cite{stepanov1976}} is that, taking the current density for a collection of charges uniformly moving at velocity $v$, $j = N (t) e v$, and assuming each charge obeys the equation of motion, $m \text{d}_t v = e E -\gamma v$, the polarization $\partial_t P = j$ obeys the second order
differential equation,
\begin{equation}
  N \frac{\partial}{\partial t} \left( \frac{1}{N} \frac{\partial P}{\partial
  t } \right) + \gamma \frac{\partial P}{\partial t} = \frac{N e^2}{m} E
  \label{eq:stepanov-model}
\end{equation}
which has the exact solution (neglecting homogeneous terms),
\begin{equation}
  P (t) = \varepsilon_0 \int_{- \infty}^t \text{d} t' \omega_p^2 (t') \int_{-
  \infty}^{t'} \text{d} t'' \text{e}^{- \gamma (t' - t'')} E (t'').
  \label{eq:P-drude}
\end{equation}
When written in terms of the Fourier expansion of the electric field, $E (t) = (2 \pi)^{- 1} \int \text{d} \omega \tilde{E} (\omega) \exp \left( - \text{i}\omega t \right)$, Eq. (\ref{eq:P-drude}) yields the following expression for the permittivity
\begin{align}
    \varepsilon_{\text{D}} (t, \omega) & = \varepsilon_{\infty} +
    \frac{\text{i}}{\text{} \omega + \text{i} \gamma} \int_{- \infty}^t
    \text{d} t' \omega_p^2 (t') \text{e}^{\text{i} \omega (t - t')}\nonumber\\
    & = \varepsilon_{\infty} - \frac{\text{$1$}}{\omega \left( \omega
    + \text{i} \gamma \right)} [\omega_p^2 (t) - \sigma_0 (t, \omega)]. \label{eq:epsilon-drude}
\end{align}
This expression equals Eq. (\ref{eq:epsilon-drude-approx}), plus an additional `memory' term $\sigma_0 = \int_{- \infty}^t \text{d} \tau \exp \left( \text{i}\omega (t - \tau) \right)  \text{d}_{\tau} \omega_p^2 (\tau)$.  Note that an important difference between Eq. (\ref{eq:epsilon-drude}) and the approximate expression (\ref{eq:epsilon-drude-approx}) is that the polarization calculated from (\ref{eq:epsilon-drude}) is always continuous, ensuring that the time modulation induces a finite current, even in the limit of a step change (see appendix \ref{sec:appendix}). If we repeatedly integrate the memory term by parts we can see that it can only be neglected when $\omega^{- n} \text{d}_{\tau}^n\omega_p^2 \ll \omega_p^2$ for all $n \geqslant 1$, i.e. to use Eq. (\ref{eq:epsilon-drude-approx}), the relative change in the plasma frequency must be small over the timescale $\omega^{- 1}$.  As Fig. \ref{fig:stepanov} shows, this `memory' term leads to significant oscillations in the permittivity when the carrier density is changed rapidly.  Despite first appearances, these oscillations do not indicate complex material dynamics,but represent the phase accumulated since the time of the modulation.

\section{Two candidate lagrangians}\label{sec:2-description}

Our first challenge is to find a field plus reservoir Lagrangian to represent something close to Stepanov's time--varying Drude model.  We now give two possible MQED Lagrangians.  Both reproduce the approximate permittivity (\ref{eq:epsilon-drude-approx}) when the relative change in the carrier density is slow, $\dot{N} / N \ll \omega$ (i.e. taking the limit shown in the upper panels of Figs. \ref{fig:approximate-drude} and \ref{fig:stepanov}).  However, as we shall see, the `obvious' modification to MQED produces a permittivity that---like the approximate expression (\ref{eq:epsilon-drude-approx})---becomes discontinuous when the carrier density is changed discontinuously.  This discontinuity implies an infinite current, which is not physical given that both the number of carriers and their energy remains finite.

\subsection{The `modulated coupling' Lagrangian}

The obvious route to the theory of MQED in a time dependent Drude
metal is to simply let the carrier density vary within the interaction
Lagrangian (\ref{eq:LInt}).  As discussed in the previous section, for the
unmodulated system, the field--reservoir coupling is given by
\begin{equation}
    \alpha (\omega) = \sqrt{\frac{2 \omega \varepsilon_0 \tmop{Im}[\varepsilon
   (\omega)]}{\pi}} = \sqrt{\frac{2 \gamma e^2 N}{\pi m^{\ast} (\omega^2 +
   \gamma^2)}} .
\end{equation}
Therefore we might expect that a time--varying carrier density would modify
the interaction Lagrangian as follows
\begin{align}
    \mathcal{L}_{\text{I}} & \rightarrow - \frac{\partial A}{\partial t}
    \int_0^{\infty} \text{d} \omega \alpha (\omega, t) X_{\omega}\nonumber\\
    & = - \frac{\partial A}{\partial t} \int_0^{\infty} \text{d} \omega
    \sqrt{\frac{2 \gamma e^2 N (t)}{\pi m^{\ast} (\omega^2 + \gamma^2)}}
    X_{\omega} .\label{eq:alpha-t}
\end{align}
Applying Lagrange's equations to the sum of (\ref{eq:LEM}--\ref{eq:LR}) and
(\ref{eq:alpha-t}), we find each oscillator in the reservoir acts as a driven, undamped simple harmonic oscillator
\begin{equation}
    \left( \frac{\partial^2}{\partial t^2} + \omega^2 \right) X_{\omega} = -\alpha (t, \omega) \frac{\partial A}{\partial t}\label{eq:eqm1}
\end{equation}
whereas the vector potential obeys the one dimensional wave equation with the
reservoir acting as a source of waves, $j = \partial_t P$,
\begin{equation}
  \begin{array}{ll}
    \left( \varepsilon_{\infty} \frac{\partial^2}{\partial t^2} - c^2
    \frac{\partial^2}{\partial x^2} \right) A & = \frac{1}{\varepsilon_0}
    \frac{\partial P}{\partial t}\\
    & \\
    & = \frac{1}{\varepsilon_0} \frac{\partial}{\partial t} \int_0^{\infty}
    \text{d} \omega \alpha (t, \omega) X_{\omega} . \label{eq:eqm2}
  \end{array}
\end{equation}
where $P$ is the polarization density.\quad The first of these coupled
equations, (\ref{eq:eqm1}) is simply that for a continuum of undamped, driven
simple harmonic oscillators.\quad This has a general solution in terms of the
retarded Green function $G_{\omega} (t - t') = \Theta (t - t') \omega^{- 1}
\sin (\omega (t - t'))$, given by
\begin{equation}
  \begin{array}{c}
    X_{\omega} = - \int_{- \infty}^t \text{d} t' G_{\omega} (t - t') \alpha
    (\omega, t') \frac{\partial A}{\partial t'}\\
    \\
    + C_{\omega}  \text{e}^{- \text{i} \omega t} + C_{\omega}^{\ast} 
    \text{e}^{\text{i} \omega t} . \label{eq:sol1}
  \end{array}
\end{equation}
where $C_{\omega}$ are arbitrary complex amplitudes that impose the $t
\rightarrow - \infty$ boundary conditions on the oscillators.\quad Inserting
the solution (\ref{eq:sol1}) into the definition of the polarization density
(\ref{eq:eqm2}) we find
\begin{align}
    P &= - \int_{- \infty}^t \text{d} t'  \int_0^{\infty} \text{d} \omega
    \alpha (t, \omega) G_{\omega} (t - t') \alpha (t', \omega) \frac{\partial A}{\partial t'}\nonumber\\
     &+ \int_0^{\infty} \text{d} \omega \alpha (t, \omega) \left( C_{\omega} 
    \text{e}^{- \text{i} \omega t} + C_{\omega}^{\ast}  \text{e}^{\text{i}
    \omega t} \right)\nonumber\\
    &= - \int_{- \infty}^t \text{d} t' \chi_{\tmop{C}} (t, t')
    \frac{\partial A}{\partial t'} + P_0 . \label{eq:polarization-naive}
\end{align}
In the final line of Eq. (\ref{eq:polarization-naive}) we have separated the
polarization into a part induced by the electric field, with the associated
susceptibility $\chi_{\tmop{C}} (t, t')$, plus the additional contribution
$P_0$ due to the undriven motion of the reservoir. Although classically we can set $P_0 = 0$, equivalent to having the reservoir at rest when $t \rightarrow- \infty$, quantum mechanically this would violate the uncertainty principle, meaning that in general we must retain $P_0$.  This so--called `noise polarization' is responsible for zero point and thermal radiation, which we return to in Sec. \ref{sec:quantum}.

\begin{figure}[h]
  \includegraphics[width=8.01575823166732cm,height=4.00787091696183cm]{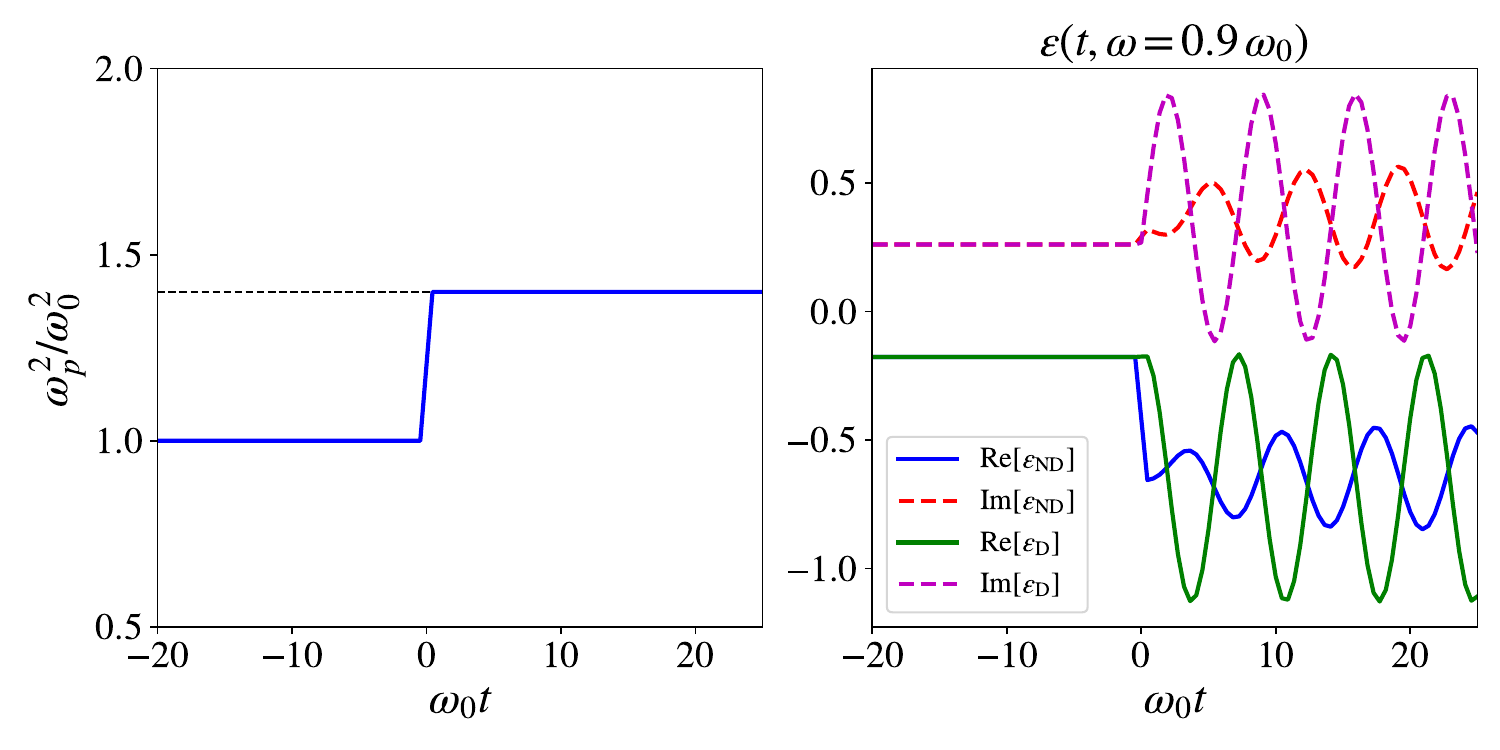}
  \caption{\textbf{Stepanov versus `modulated coupling' Lagrangian:} (a) the
  plasma frequency changes abruptly as $\omega_p^2 (t) = \omega_0^2 + \delta
  \omega_p^2 \frac{1}{2} [1 + \tanh (t / t_r)]$ with $\omega_0 t_r = 0.05$ and $\delta \omega_p^2 = 0.4 \omega_0$. (b) Comparison between the
  permittivities $\varepsilon (\omega, t)$ given by the Stepanov model
  (\ref{eq:epsilon-drude}) and calculated from the `modulated coupling'
  Lagrangian (\ref{eq:epsilon-L1}), for the fixed frequency $\omega = 0.9
  \omega_0$.  Note that the permittivity (\ref{eq:epsilon-L1}) changes
  extremely rapidly around $t = 0$, whereas (\ref{eq:epsilon-drude}) remains
  continuous.  Both permittivity functions exhibit the oscillations noted earlier in Fig. \ref{fig:stepanov}. \label{fig:compare-epsilon}}
\end{figure}

The relative permittivity is given in terms of the above two--time
susceptibility as \ $\varepsilon_{\rm C} (t, t') = \varepsilon_{\infty} + \chi_{\rm C} (t,
t')$.\quad To compare this with the earlier results for the time--dependent
Drude model (\ref{eq:epsilon-drude}), we perform a Fourier transform of the
permittivity in the second time argument, $t'$ giving,
\begin{equation}
  \varepsilon_{\text{C}} (t, \omega) = \varepsilon_{\infty} - \frac{1}{\omega
  \left( \omega + \text{i} \gamma \right)} [\omega_p^2 (t) - \sigma_1 (t,
  \omega)] \label{eq:epsilon-L1}
\end{equation}
where the `memory' term $\sigma_1$ is here defined as
\begin{multline}
    \sigma_1 = \frac{\text{i} \omega_p (t)}{\gamma} \int_{- \infty}^t \text{d}t' \frac{\text{d} \omega_p (t')}{\text{d} t'} \text{e}^{\text{i} \omega (t- t')}\\
    \times \left( \omega \left( \text{e}^{- \gamma (t - t')} - 1\right) - \text{i} \gamma \right) .
\end{multline}
The effective permittivity (\ref{eq:epsilon-L1}) derived from the Lagrangian
of MQED (\ref{eq:eqm1}--\ref{eq:eqm2}) has the same form as the
earlier time dependent Drude model (\ref{eq:epsilon-drude}), with the same
limiting form (\ref{eq:epsilon-drude-approx}), when the carrier density
changes slowly.  However there are some extremely important differences
between this permittivity and the model given in Ref. {\cite{stepanov1976}}
(i.e. Eq. (\ref{eq:epsilon-drude})).  Most importantly the permittivity in
Eq. (\ref{eq:epsilon-L1}) is discontinuous when the carrier density undergoes a temporal discontinuity (see Fig. \ref{fig:compare-epsilon}).  By
implication this leads to a discontinuous polarization, and thus an \emph{infinite} current density, $\partial_t P$.  Clearly this does not reflect the expected behaviour from a sudden but \emph{finite} increase in carrier density.  The singular current comes directly from our introduction of the time dependent coupling in Eq. (\ref{eq:alpha-t}): while the reservoir equation of motion (\ref{eq:eqm1}) ensures continuity of both $X_{\omega}$ and $\partial_t X_{\omega}$, neither the polarization $P = \int \text{d} \omega \alpha (\omega, t) X_{\omega}$ nor its time derivative inherit this continuity when $\alpha$ is changed abruptly.

Just as importantly, the `noise polarization' shows the same discontinuity as a function of time,
\begin{equation}
  \begin{array}{cl}
    P_0 & = \int_0^{\infty} \text{d} \omega \alpha (t, \omega) \left(
    C_{\omega}  \text{e}^{- \text{i} \omega t} + C_{\omega}^{\ast} 
    \text{e}^{\text{i} \omega t} \right)\\
    & \\
    & = \omega_p (t) \int_0^{\infty} \text{d} \omega \sqrt{\frac{2 \gamma
    \varepsilon_0}{\pi (\omega^2 + \gamma^2)}} \left( C_{\omega}  \text{e}^{-
    \text{i} \omega t} + C_{\omega}^{\ast}  \text{e}^{\text{i} \omega t}
    \right), \label{eq:noise-coupling}
  \end{array}
\end{equation}
meaning that, when we quantize the time--varying Drude model with the
modulated coupling (\ref{eq:alpha-t}), we'd predict arbitrarily large
fluctuating thermal currents, constrained only by the timescale over which we can modulate the carrier density (as shown in Fig. \ref{fig:noise}).

\subsection{The `modulated reservoir' Lagrangian}

It thus appears that introducing a time--modulation within the field--reservoir coupling in MQED does not accurately capture the effect of changing the carrier density in a metal.  As we've seen, this leads to spuriously large transient currents.

We now remove the carrier density from the field--reservoir coupling
\begin{equation}
  \alpha (\omega) \rightarrow \sqrt{\frac{2 \gamma e^2}{\pi m^{\ast} (\omega^2
  + \gamma^2)}} \label{eq:coupling-2}
\end{equation}
and instead introduce it as a scaling of the reservoir Lagrangian,
\begin{equation}
  \mathcal{L}_{\text{R}} \rightarrow \frac{1}{2} \int_0^{\infty} \text{d}
  \omega \frac{1}{N (t)} \left[ \left( \frac{\partial X_{\omega}}{\partial t}
  \right)^2 - \omega^2 X_{\omega}^2 \right] . \label{eq:LR-2}
\end{equation}
This makes a modulation of the carrier density equivalent to a change in the
mass of every oscillator.  For e.g. a large concentration of carriers this
effective mass is small, indicating that then---as expected---it is easy to
polarize the material.

Applying Lagrange's equations to the sum of (\ref{eq:LEM}), (\ref{eq:LInt}),
and (\ref{eq:LR-2}), the electromagnetic wave equation takes the same form as earlier, given in Eq. (\ref{eq:eqm2}).  The dynamics of the reservoir,
however, is now governed by
\begin{equation}
    N \frac{\partial }{\partial t} \left( \frac{1}{N} \frac{\partial
    X_{\omega}}{\partial t} \right) + \omega^2 X_{\omega} = - N (t) \alpha
    (\omega) \frac{\partial A}{\partial t}\label{eq:res-eqm2}
\end{equation}
which is very similar to the differential equation for the polarization
(\ref{eq:stepanov-model}).  Indeed, multiplying Eq. (\ref{eq:res-eqm2}) by
the coupling constant $\alpha (\omega)$ and integrating over $\omega$, Eq.
(\ref{eq:res-eqm2}) becomes
\begin{equation}
    N \frac{\partial }{\partial t} \left( \frac{1}{N} \frac{\partial
    P}{\partial t} \right) + \int_0^{\infty} \text{d} \omega \omega^2 \alpha
    (\omega) X_{\omega} = - \frac{e^2 N (t)}{m^{\ast}} \frac{\partial
    A}{\partial t},
    \label{eq:modeqn}
\end{equation}
which besides the damping term, is the same as the differential equation
governing the time--varying Drude model (\ref{eq:stepanov-model}).  Indeed, given that the plasma frequency and the damping arise from coupling to the same system (here the reservoir) it is questionable whether the damping
should be assumed constant as it is in Eq. (\ref{eq:stepanov-model}) and Ref.~\cite{stepanov1976}.

For a general variation of the carrier density, it is difficult to solve
(\ref{eq:res-eqm2}).  We take the special case where there is an abrupt
change at $t = 0$: $N (t) = N_0 \Theta (- t) + N_1 \Theta (t)$ (the plasma
frequency accordingly changing from $\omega_0$ to $\omega_1$ at $t = 0$).  In this case the solution to Eq. (\ref{eq:res-eqm2}) is
\begin{multline}
    X_{\omega} = - \int_{- \infty}^t \text{d} t' \frac{\alpha
    (\omega)}{\omega} \frac{\partial A}{\partial t'} [\sin (\omega (t - t')) N
    (t') \nobracket\\[5pt]
    \nobracket + \Theta (t) \Theta (- t') N_0 \Delta \sin (\omega t) \cos
    (\omega t')] + X_{0 \omega} .\label{eq:res-sol}
\end{multline}
where $\Delta = 1 - N_1 / N_0$.  The final quantity $X_{0 \omega}$ is the
homogeneous solution to Eq. (\ref{eq:res-eqm2}), defined---as in the previous section---in terms of a set of complex amplitudes $C_{\omega}$,
\begin{align}
    X_{0 \omega} &= C_{\omega} \left[ \text{e}^{- \text{i} \omega t} +
    \text{i} \Delta \Theta (t) \sin (\omega t) \right] + \text{c} . \text{c}
    .\nonumber\\[5pt]
    & = C_{\omega} \xi_{\omega} (t) + C_{\omega}^{\ast} \xi_{\omega}^{\ast}
    (t) \label{eq:homsol}
\end{align}
Inserting the solution (\ref{eq:res-sol}) for the reservoir dynamics into the definition of the polarization and performing a Fourier transform, we find the effective permittivity is ag given by,
\begin{equation}
  \varepsilon_{\rm R} (t, \omega) = \varepsilon_{\infty} - \frac{1}{\omega \left(
  \omega + \text{i} \gamma \right)} [\omega_p^2 (t) - \sigma_2 (t, \omega)]
  \label{eq:epsilon-modified}
\end{equation}
where the `memory' term is now given by,
\begin{equation}
    \sigma_2 (t, \omega) = \Theta (t) \frac{\omega_{p 0}^2 \Delta}{1 -
    \frac{\text{i} \gamma}{\omega}} \left[ \frac{\text{i} \gamma}{\omega}
    \text{e}^{2 \text{i} \omega t} - \text{e}^{(i \omega - \gamma) t}
    \right]
    \label{eq:memory-2}
\end{equation}
When $t \rightarrow 0^+$ the above expression reduces to $\sigma_2 =
\omega_1^2 - \omega_0^2$ showing that the permittivity
(\ref{eq:epsilon-modified}) is continuous across the jump in carrier density
at $t = 0$.\quad Note that, although the `memory' term (\ref{eq:memory-2})
\tmtextit{appears} to have a pole in the upper half of the complex frequency
plane at $\omega = + i \gamma$, which would violate the Kramers--Kronig
relations {\cite{solis2020}}, the residue of this pole is zero.

\begin{figure}[h]
  \includegraphics[width=8.01575823166732cm,height=4.00787091696183cm]{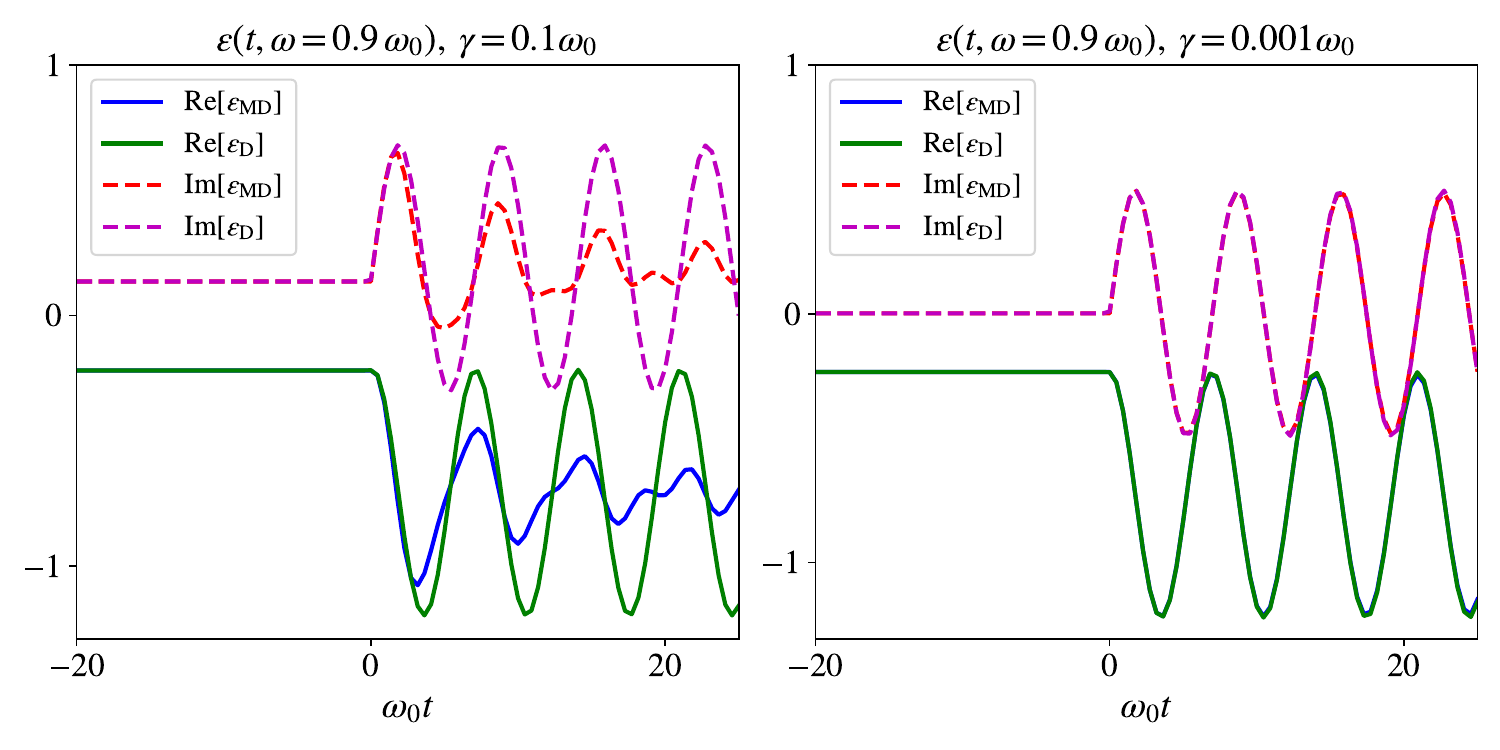}
  \caption{\textbf{Stepanov versus `modulated reservoir' Lagrangian:}
  Comparison between the permittivity (\ref{eq:epsilon-drude}) derived from
  the Stepanov model and that derived from the ``mature Drude'' Lagranigian
  (\ref{eq:epsilon-modified}), for the modulation shown in Fig.
  \ref{fig:compare-epsilon}a and the fixed frequency $\omega = 0.9
  \omega_0$.  (a) The case of ``large'' damping, $\gamma = 0.1 \omega_0$,
  showing that $\varepsilon_R$ and $\varepsilon_{\text{D}}$ oscillate in
  phase, with the reservoir approach yielding an additional harmonic and an
  exponential decay. (b) In the limit of ``small'' damping, $\gamma = 0.001
  \omega_0$, the two permittivity functions, $\varepsilon_{\text{D}}$ and
  $\varepsilon_R$ become equal.\label{fig:epmvseps}}
\end{figure}

Figure \ref{fig:epmvseps} shows a comparison between the permittivity,
$\varepsilon (t, \omega)$ computed from the Stepanov model (\ref{eq:epsilon-drude}) and that we just derived from the Lagrangian density with the modified reservoir (\ref{eq:LR-2}).  Unlike the earlier
permittivity (\ref{eq:epsilon-L1}), derived from modulating the
field--reservoir coupling, $\epsilon_C$ is continuous at $t = 0$ and becomes
equal to the permittivity of the Stepanov model in the limit of small damping, $\gamma / \omega \ll 1$.  In this small--damping limit we have therefore found a Lagrangian description of a time--varying Drude metal that we can quantize.

\begin{figure}[h]
  \includegraphics[width=5.3426800472255cm,height=4.01833267742359cm]{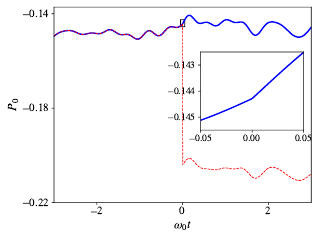}
  \caption{\textbf{Visualizing noise polarization dynamics:}  Taking $N
  = 3500$ frequency points evenly spaced between $\omega = 0$ and $15
  \omega_0$, the magnitude and phase of the complex amplitudes $C_{\omega}$
  were chosen randomly to compute the classical noise polarization arising
  from the `modulated coupling' (\ref{eq:noise-coupling}) (red dashed) and
  `modulated reservoir' (\ref{eq:noise-drudem}) (blue solid) approaches, for a large, abrupt change $\omega_p^2 = \omega_0^2 \rightarrow 2 \omega_0^2$ in plasma frequency at $t = 0$ ($\gamma = 0.001 \omega_0$).\quad While the
  `modulated coupling' Lagrangian predicts a discontinuous noise polarization, the `modulated reservoir' approach predicts a discontinuous time derivative (shown inset). \label{fig:noise}}
\end{figure}

As shown above, the homogeneous solutions to Eq. (\ref{eq:res-eqm2}) are given by (\ref{eq:homsol}), which are not simple complex exponentials.  Substituting these expressions into the definition of the noise polarization, $P_0 =\int_0^{\infty} \text{d} \omega \alpha (\omega) X_{0 \omega}$, we find this equals
\begin{align}
    P_0 & = \int_0^{\infty} \text{d} \omega \alpha (\omega) [C_{\omega}
    \xi_{\omega} (t) + C_{\omega}^{\ast} \xi_{\omega}^{\ast} (t)]\nonumber\\
    & = \int_0^{\infty} \text{d} \omega \sqrt{\frac{\frac{2 \gamma e^2}{\pi
    m^{\ast}}}{\omega^2 + \gamma^2}} [C_{\omega} \xi_{\omega} (t) +
    C_{\omega}^{\ast} \xi_{\omega}^{\ast} (t)]\label{eq:noise-drudem}
\end{align}
Unlike the noise polarization (\ref{eq:noise-coupling}), which is
discontinuous in time when the plasma frequency is changed abruptly, we now
have a noise polarization (\ref{eq:noise-drudem}) that is continuous, although its first derivative is not (see Fig. \ref{fig:noise}).  This rapid change in the derivative of the noise current $\partial_t j = \partial_t^2 P_0$ will lead to a `flash' of radiation at $t = 0$.

Note that although the results of this section are only valid for a step
change in the carrier density, we can approximate any continuous change in
density as a series of such steps, using e.g. a transfer matrix to generally
compute the time evolution of the oscillator amplitudes $X_{\omega}$.

\section{Quantization}\label{sec:quantum}

We have thus found a Lagrangian density that---through its equation of
motion---leads to Maxwell's equations with a permittivity that tends to
Stepanov's expression (\ref{eq:epsilon-drude}) in the limit of small
damping.  Although we make no claim that this particular model generally
represents the dynamics of a real metal, it is clear the predictions of
MQED very much depend on exactly how the time modulation is
included within the Lagrangian.  While a modulated field--reservoir
coupling (\ref{eq:alpha-t}) can yield unphysically large polarization
currents, modulating the effective mass of the reservoir oscillators
(\ref{eq:LR-2}) does not suffer this defect.  We shall now compare the predictions of the quantum counterparts of these theories, giving the counterpart of the fluctuation--dissipation therorem for the two Lagrangians described above.

The first step is to derive the Hamiltonian density, written in terms of a set of field variables and their conjugate momenta. For the reservoir variables the canonical momentum density, $\Pi_{\omega} = \delta \mathcal{L} / \delta \dot{X}_{\omega}$ is given by, in the cases of ``modulated coupling'' (C) and ``modulated reservoir'' (R) by,
\begin{equation}
  \Pi_{\omega} = \begin{cases}
  \frac{\partial X_{\omega}}{\partial t} & \text{(C)}\\[5pt]
  \frac{1}{N (t)} \frac{\partial X_{\omega}}{\partial t} & \text{(R)}.
  \end{cases}
\end{equation}
and similarly for the vector potential of the EM field, where the canonical
momentum density, $\Pi_A = \delta \mathcal{L} / \delta \dot{A}$ equals
\begin{equation}
  \Pi_A =   \begin{cases}
  \varepsilon_0 \varepsilon_{\infty} \frac{\partial A}{\partial t} -
  \int_0^{\infty} \text{d} \omega \alpha (\omega, t) X_{\omega}&\text{(C)}\\[5pt]
  \varepsilon_0 \varepsilon_{\infty} \frac{\partial A}{\partial t} -
  \int_0^{\infty} \text{d} \omega \alpha (\omega) X_{\omega} & \text{(R)}
  \end{cases}
\end{equation}
where the field--reservoir coupling is defined in Eqns. (\ref{eq:alpha-t}) and (\ref{eq:coupling-2}) respectively.  Forming the classical Hamiltonian
density via the usual expression and separating the result into three terms,
\begin{align}
    \mathcal{H} & = \Pi_A \partial_t A + \int_0^{\infty} \text{d} \omega
    \Pi_{\omega} \partial_t X_{\omega} - \mathcal{L}\\[5pt]
    & = \mathcal{H}_{\text{E}} + \mathcal{H}_{\text{B}} +
    \mathcal{H}_{\text{R}}
\end{align}
we find, in the two cases described above, we have the reservoir energy,
\begin{equation}
  \mathcal{H}_{\text{R}} = \begin{cases}
  \frac{1}{2} \int_0^{\infty} \text{d} \omega (\Pi_{\omega}^2 + \omega^2
  X_{\omega}^2) & \text{(C)}\\[5pt]
  \frac{1}{2} \int_0^{\infty} \text{d} \omega \left( N (t) \Pi_{\omega}^2 +
  \frac{\omega^2}{N (t)} X_{\omega}^2 \right) & \text{(R)}
  \end{cases}
\end{equation}
and the energy stored in the electric field,
\begin{equation}
    \mathcal{H}_{\text{E}} = \frac{1}{2 \varepsilon_0 \varepsilon_{\infty}}\times
    \begin{cases}
    \left( \Pi_A + \int_0^{\infty} \text{d} \omega \alpha (\omega, t)
      X_{\omega} \right)^2 & \text{(C)}\\[5pt]
      \left( \Pi_A + \int_0^{\infty} \text{d} \omega \alpha (\omega)
      X_{\omega} \right)^2 & \text{(R)}
    \end{cases}
\end{equation}
In both cases the magnetic energy equals $\mathcal{H}_{\text{B}} = \frac{1}{2
\mu_0} (\partial_x A)^2$.  Promoting these classical field and momentum
variables to operators with the equal--time commutation relations,
$[\hat{X}_{\omega} (x, t), \hat{\Pi}_{\omega'} (x', t)] = \text{i} \hbar
\delta (\omega - \omega') \delta (x - x')$ and $[\hat{A} (x, t), \hat{\Pi}_A
(x', t)] = i \hbar \delta (x - x')$, the quantum mechanical Hamiltonian is
then given by $\hat{H} = \int \hat{\mathcal{H}}  \text{d} x$.
\begin{figure}[h]
  \includegraphics[width=8.01575823166732cm,height=4.00787091696183cm]{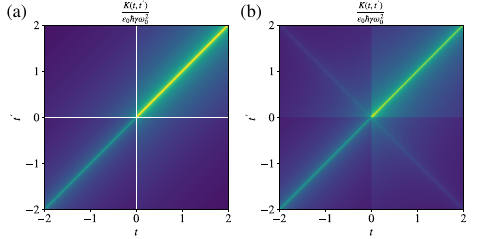}
  \caption{\textbf{Noise current autocorrelation:}  Normalized in units
  of $\varepsilon_0 \hbar \gamma \omega_0^2$ and for the parameters $N_1 = 1.4 N_0$ and $\gamma = 0.1$. (a) Computed for the `modulated coupling' theory (\ref{eq:autocorrelation-C}), the white lines indicating where the
  autocorrelation is singular, and (b) for the `modulated reservoir' theory
  (\ref{eq:autocorrelation-R}), which is discontinuous but finite. Note
  the presence of correlations along $t = - t'$.\label{fig:jj-correlation}}
\end{figure}
Due to the time dependence of the material, it might be questioned whether we
can use this Hamiltonian operator to usefully compute thermal averages or vacuum radiation, given the lack of thermal equilibrium.  We can use the
Heisenberg picture to sidestep this difficulty, working with the states $|
\psi \rangle \nobracket$ of the system defined at $t = - \infty$ and placing
all the time evolution within the expressions for the operators, any operator
$\hat{O}$ obeying $\text{d}_t \hat{O} = \frac{\text{i}}{\hbar} [\hat{H},
\hat{O}] + \partial_t \hat{O}$.  Evaluating these operator equations of
motion, we find they are formally identical to the classical equations of
motion e.g. (\ref{eq:eqm1}), (\ref{eq:eqm2}) and (\ref{eq:res-eqm2}), as
expected due to the quadratic form of the Hamiltonian.

This formal equivalence between classical and operator equations of motion
means we can solve the quantum mechanical evolution through simply promoting
our classical solutions to operator expressions.\quad For instance, the
homogeneous solutions (\ref{eq:sol1}) and (\ref{eq:homsol}) to the operator
dynamics of the reservoir are,
\begin{equation}
  \hat{X}_{0, \omega} = \begin{cases}
  \sqrt{\frac{\hbar}{2 \omega}} \left( \hat{C}_{\omega} \text{e}^{- \text{i}
    \omega t} + \hat{C}_{\omega}^{^{\dag}} \text{e}^{\text{i} \omega t}
    \right) & \text{(C)}\\[5pt]
    \sqrt{\frac{\hbar N_0}{2 \omega}} (\hat{C}_{\omega} \xi_{\omega} (t) +
    \hat{C}_{\omega}^{^{\dag}} \xi_{\omega}^{\ast} (t)) & \text{(R)}
  \end{cases}\label{eq:reservoir-ops}
  % \left\{ \begin{array}{lc}
  %   \sqrt{\frac{\hbar}{2 \omega}} \left( \hat{C}_{\omega} \text{e}^{- \text{i}
  %   \omega t} + \hat{C}_{\omega}^{^{\dag}} \text{e}^{\text{i} \omega t}
  %   \right) & \text{(C)}\\
  %   & \\
  %   \sqrt{\frac{\hbar N_0}{2 \omega}} (\hat{C}_{\omega} \xi_{\omega} (t) +
  %   \hat{C}_{\omega}^{^{\dag}} \xi_{\omega}^{\ast} (t)) & \text{(R)}
  % \end{array} \right. \label{eq:reservoir-ops}
\end{equation}
where $\xi_{\omega}$ is defined in Eq. (\ref{eq:homsol}) and the prefactors of $\sqrt{\hbar / 2 \omega}$ and $\sqrt{\hbar N_0 / 2 \omega}$ are chosen to
enforce the canonical commutation relations, $[\hat{X}_{0, \omega} (x, t),
\hat{\Pi}_{0, \omega'} (x', t)] = \text{i} \hbar \delta (x - x') \delta
(\omega - \omega')$.  We have promoted the complex classical amplitudes
$C_{\omega}$ to bosonic operators satisfying the continuum commutation relations
\begin{equation}
  [\hat{C}_{\omega} (x), \hat{C}_{\omega'}^{^{\dag}} (x')] = \delta (\omega -
  \omega') \delta (x - x') .
\end{equation}
with $\hat{C}_{\omega}^{^{\dag}}$ and $\hat{C}_{\omega}$ respectively
interpreted as `polariton' creation and annihilation operators.  As in the classical discussion above, the noise polarization arising from our two Lagrangians---defined in (\ref{eq:polarization-naive}) and (\ref{eq:noise-drudem})---are given as integrals over the homogenous solutions
to the reservoir dynamics,
\begin{equation}
    \hat{P}_0 = 
    \begin{cases}
      \int_0^{\infty} \text{d} \omega \sqrt{\frac{\varepsilon_0 \hbar \gamma
      \omega_p^2 (t)}{\pi \omega (\omega^2 + \gamma^2)}}  \hat{C}_{\omega} 
      \text{e}^{- \text{i} \omega t} + \text{h.c.} & \text{(C)}\\[5pt]
      \int_0^{\infty} \text{d} \omega \sqrt{\frac{\varepsilon_0 \hbar \gamma
      \omega_0^2}{\pi \omega (\omega^2 + \gamma^2)}}  \hat{C}_{\omega}
      \xi_{\omega} (t) + \text{h.c.} & \text{(R)}
    \end{cases}\label{eq:P0-op}
\end{equation}
where we have replaced the coupling functions $\alpha (\omega)$ and $\alpha
(\omega, t)$ with their definitions in (\ref{eq:alpha-t}) and
(\ref{eq:coupling-2}).\quad Remember that to compare the two operators in Eq.
(\ref{eq:P0-op}) we must consider an abrupt change in plasma frequency, from
$\omega_0$ to $\omega_1$ at $t = 0$.

In terms of the noise polarization (\ref{eq:P0-op}), the noise current is
given by, $\hat{j}_0 = \partial \hat{P}_0 / \partial t$.  This noise
current is the source of the quantum fluctuations in the electromagnetic field indicated in Fig. \ref{fig:schematic}, usually predicted from the
fluctuation--dissipation theorem.  As a concrete example we consider these
fluctuating currents when the system is initially in its vacuum state $|0\rangle$, defined as the zero polariton state,
\begin{equation}
  \hat{C}_{\omega} (x) | 0 \rangle = 0 \nobracket \label{eq:vacuum-state}.
\end{equation}
To understand the noise currents within our two models, here we compare their
vacuum current--current correlation functions,
\begin{multline}
    \frac{1}{2} \langle 0 | \hat{j}_0 (x, t) \hat{j}_0 (x', t') + \hat{j}_0
    (x', t') \hat{j}_0 (x, t) | 0 \rangle\\[5pt]
    = \delta (x - x') K (t, t'),
\end{multline}
where we define $K (t, t')$ as the time autocorrelation of the noise current at a fixed point in space.  Computing this autocorrelation function for the
`modulated coupling' theory, using both Eqns. (\ref{eq:P0-op}) and (\ref{eq:vacuum-state}) we have the autocorrelation,
\begin{multline}
    K (t, t') = \int_0^{\infty} \frac{\text{d} \omega}{2 \pi}
    \frac{\varepsilon_0 \hbar \gamma}{\omega (\omega^2 + \gamma^2)} [\delta
    (t) \omega_0 \Delta + i \omega \omega_p (t)]\\[5pt]
    \times [\delta (t') \omega_0 \Delta - i \omega \omega_p (t')] \text{e}^{-
    \text{i} \omega (t - t')} + \text{c.c.} \qquad
    \text{$\label{eq:autocorrelation-C}$(C)}
\end{multline}
whereas, for the `modulated reservoir' theory the autocorrelation is,
\begin{multline}
    K (t, t') = \int_0^{\infty} \frac{\text{d} \omega}{2 \pi}
    \frac{\varepsilon_0 \hbar \gamma \omega \omega_0^2}{\omega^2 + \gamma^2}
    \left( \text{e}^{- \text{i} \omega t} - \Delta \Theta (t) \cos (\omega t)
    \right)\\[5pt]
    \tmcolor{white}{} \times \left( \text{e}^{\text{i} \omega t'} - \Delta
    \Theta (t') \cos (\omega t') \right) + \text{c.c.} \qquad
    \label{eq:autocorrelation-R} \text{(R)}
\end{multline}
Taking the limit of an unmodulated carrier density, $N_1 = N_0$, both
(\ref{eq:autocorrelation-C}) and (\ref{eq:autocorrelation-R}) reduce to the
same expression
\begin{align}
    K (t - t') & = \varepsilon_0 \hbar \int_0^{\infty} \frac{\text{d}
    \omega}{2 \pi} \frac{\gamma \omega \omega_p^2}{\omega^2 + \gamma^2}
    \text{e}^{- \text{i} \omega (t - t')} + \text{c.c.}\nonumber\\[5pt]
    & = \varepsilon_0 \hbar \int_0^{\infty} \frac{\text{d} \omega}{2 \pi}
    \omega^2 \tmop{Im} [\varepsilon (\omega)] \text{e}^{- \text{i} \omega (t -t')} + \text{c.c.} \label{eq:fluc-diss}
\end{align}
the Fourier transform of which equals
\begin{equation}
  \begin{array}{cl}
    \tilde{K} (\omega) & = \varepsilon_0 \hbar \omega^2 \tmop{sign} (\omega)
    \tmop{Im} [\varepsilon (\omega)] . \label{eq:corr0}
  \end{array}
\end{equation}
Equation (\ref{eq:corr0}) is the zero temperature limit of the
fluctuation--dissipation theorem {\cite{kubo1966,volume9}}, a well--known
relation between the spectrum of the noise current fluctuations (here
$\tilde{K}$) and the imaginary part of the susceptibility.

Fig. \ref{fig:jj-correlation} compares the two current autocorrelation
functions, (\ref{eq:autocorrelation-C}) and (\ref{eq:autocorrelation-R})
derived above.  As anticipated from our classical analysis, in the case of
the `modulated coupling' theory, Fig. \ref{fig:jj-correlation}a shows that the autocorrelation of the current is singular along the lines $t = 0$ and $t' = 0$, due to the abrupt change in the carrier density.  Moreover, as a
simple analysis shows, away from these lines, where we can neglect the delta
functions in Eq. (\ref{eq:autocorrelation-C}), the correlation is simply that
given by the fluctuation--dissipation theorem (\ref{eq:fluc-diss}), scaled by
the product of the plasma frequencies at the two times, $t$ and $t'$, divided
by $\omega_0^2$.

By contrast, the `modulated reservoir' theory yields a non--singular
autocorrelation function, albeit one that is discontinuous across the lines $t= 0$ and $t' = 0$.  An important feature shown in Fig. \ref{fig:jj-correlation}b is that this theory develops additional correlations
in the noise current along the lines $t = \pm t'$.  These arise because in this case we are modulating the properties of the reservoir, which leads to
the temporal analogue of reflection in the oscillator dynamics.  Modulating the field--reservoir coupling cannot ever lead to such correlations.

\section{Summary and conclusions}

Although MQED is a well established tool for predicting quantum
behaviour in real--world materials, here we have shown that it is non--trivial to extend this theory to the case of time--varying media.

To include the effects of dispersion and dissipation, MQED makes
use of a fictitious reservoir of harmonic oscillators, the linear response of
which perfectly matches that of the real medium, specifying the coupling
between the field and the reservoir in terms of the imaginary part of the
susceptibilities.  When we quantize this theory we are thus quantizing
electromagnetism coupled to a system with a linear response that exactly
matches that of the material, as measured in the laboratory.  We have
shown that the usual simple relationship between the system--reservoir coupling and the permittivity is broken when the material response is modulated in time.

Using the example of a Drude model with a time--varying carrier density we
have shown that a naive replacement of $\tmop{Im} [\varepsilon (\omega)]
\rightarrow \tmop{Im} [\varepsilon (\omega, t)]$ in the standard macroscopic
QED theory leads to a polarization of the material that can change as rapidly
as we can modulate the carrier density, the polarization current becoming
singular in the limit of a discontinuous change (which non--physical, given that both the total charge and energy of the reservoir remains finite).  As an alternative we considered a theory where the change in carrier density is incorporated through modifying the reservoir dynamics.  This alternative model leads to a permittivity that reproduces the classical time varying Drude model of Stepanov~\cite{stepanov1976} in the limit of small damping, with the polarization current remaining finite.

One of the central predictions of MQED is the presence of noise currents within dispersive, dissipative media.  The presence of these currents is necessary to retain thermal equilibrium, and they are missed in both classical theories, and quantum theories that neglect dispersion.  Again, using the example of a time--modulated Drude model with an abrupt change in the carrier density, we have shown that the discontinuous polarization of the ``modulated coupling'' theory of MQED yields an infinite noise current.  Meanwhile, our alternative ``modulated reservoir'' theory predicts a finite noise current, plus additional correlations that arise from `time reflection' within the reservoir dynamics, due to the modulated parameters.

We have shown that a theory of MQED with modulated reservoir
parameters more closely models the dynamics of a metal with a modulated
carrier density, compared with a naive modification of the standard
theory.  This `modulated reservoir' theory leads to new phenomena, such as additional correlations within the noise currents, which although difficult to observe in the vacuum state, may be observable in higher temperature thermal correlations.  More generally it also remains an interesting question as to how to incorporate the precise linear response of any particular time--varying material within a Lagrangian description.

\begin{acknowledgments*}
SARH and RKB acknowledge financial support from the Royal Society and TATA
(RPG-2016-186), and SARH acknowledges support from the EPSRC through the
program grant ``Next generation metamaterials: exploiting four dimensions''
(EP/Y015673/1).\quad SARH acknowledges useful conversations with J. B. Pendry, E. Hendry, A. Ganfornina, J. Echave--Sustaeta, and C. Hooper.
\end{acknowledgments*}

\subsection{Appendix: continuity of the permittivty}\label{sec:appendix}

\begin{figure}[h]
  \includegraphics[width=8.01575823166732cm,height=4.00787091696183cm]{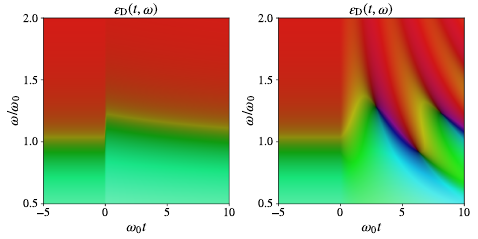}
  \caption{\textbf{Continuity of the Stepanov permittivity:} Comparison
  between (left) the approximate Drude model (\ref{eq:epsilon-drude-approx})
  and (right) the Stepanov model (\ref{eq:epsilon-drude}) for an abrupt change in carrier density $\omega_0 t_r = 0.05$ and $\omega_0 t_d = 10$ (see Fig. \ref{fig:approximate-drude}).\label{fig:continuity}}
\end{figure}

Here we give an example illustrating the continuity of the Stepanov
permittivity (\ref{eq:epsilon-drude}) in the case of an abrupt change in the
carrier density. Fig. \ref{fig:continuity} compares the complex functions
(\ref{eq:epsilon-drude-approx}) and (\ref{eq:epsilon-drude}), showing that the approximate permittivity tends towards being discontinuous.

\end{document}